\documentclass[conference]{IEEEtran}
\IEEEoverridecommandlockouts
\usepackage{amsmath,amssymb,amsfonts}
\usepackage{graphicx}
\usepackage{hhline}
\usepackage{colortbl}
\usepackage{multicol}
\usepackage[usenames,dvipsnames]{xcolor}
\usepackage[obeyspaces,spaces]{url}
\usepackage{multicol}
\usepackage{xspace}
\usepackage{booktabs}
\usepackage{cleveref}
\usepackage{multirow}
\usepackage[ruled,vlined,linesnumbered]{algorithm2e}
\def\BibTeX{{\rm B\kern-.05em{\sc i\kern-.025em b}\kern-.08em
    T\kern-.1667em\lower.7ex\hbox{E}\kern-.125emX}}
\begin{document}

\title{Online and Real-time Object Tracking Algorithm with Extremely Small Matrices}

\author{\IEEEauthorblockN{Jesmin Jahan Tithi\IEEEauthorrefmark{1}, Sriram
    Aananthakrishnan\IEEEauthorrefmark{2} and Fabrizio
    Petrini\IEEEauthorrefmark{1}}
  \IEEEauthorblockA{
    \IEEEauthorrefmark{1}Intel Parallel Computing Labs,
    \IEEEauthorrefmark{2}Intel Data Center Group
    \\
    Email: \{jesmin.jahan.tithi,
    sriram.aananthakrishnan,
    fabrizio.petrini\}@intel.com}
}
\maketitle

\begin{abstract}
Online and Real-time Object Tracking is an interesting workload that can be used to track objects (e.g., car, human, animal) in a series of video sequences in real-time. For simple object tracking on edge devices, the output of object tracking could be as simple as drawing a bounding box around a detected object and in some cases, the input matrices used in such computation are quite small (e.g., 4x7, 3x3, 5x5, etc). As a result, the amount of actual work is low. Therefore, a typical multi-threading based parallelization technique can not accelerate the tracking application; instead, a throughput based parallelization technique where each thread operates on independent video sequences is more rewarding. In this paper, we share our experience in parallelizing a Simple Online and Real-time Tracking (SORT) application on shared-memory multicores.
\end{abstract}


\section{Introduction}
Online and Real-time object tracking can be used to track objects in a series of video sequences in real-time and has applications in object detection systems such as autonomous vehicles, security cameras, other surveillance, and robotics. Because of the real-time and online nature of the task, it is often latency-sensitive. For simple object tracking (especially, on edge devices, mobile phones, cameras), the output could be as simple as drawing bounding boxes around the target objects. For this task, the input matrices are often very small and as a result, a typical multi-threading based parallelization technique to accelerate these applications is not scalable. In fact, a well-optimized serial code on a high-end x86 CPU would perform better than a multi-threaded code due to scheduling and sychronization overheads with multi-threading.

In this paper, we share our experience of parallelizing a Simple Online and Real-time Tracking (SORT) workload \cite{SORT, SORT_paper} for shared-memory multicores. The SORT \cite{SORT, SORT_paper} application was originally written in python and used many python libraries such as filterpy \cite{filterpy} and linear assignment \cite{scikit-learn} libraries, parallel matrix libraries and BLAS routines for matrix operations. We implemented a C version of it and our sequential C version is $44\times$ to $100\times$ faster than the parallel python code. We also parallelized the code using OpenMP and we shared our insights and lessons learned during that work which we found quite challenging. 

Our contributions in this paper are:
\begin{itemize}
\item We present a very thorough analysis of the Simple Online and Real-time Tracking (SORT) workload including the runtime analysis of the original python code and our C code for SORT. 
\item We re-implement the SORT code in $C$ which is over $44$ to $100\times$ times faster than the original parallel python implementation.
\item We parallelize the code using openMP and show scalability analysis using three different types of scaling methods: strong scaling, weak scaling, and throughput scaling and conclude that for this workload and with the type of dataset we used, weak-scaling or throughput scaling is more rewarding.
\end{itemize} 

\section{Background}

The main part of the SORT algorithm consists of a combination of the Kalman filter \cite{KalmanFilter0} and the Hungarian algorithm \cite{kuhn1955hungarian} (e.g, optimal cost assignment problem), to identify and track target objects in a video frame. SORT analyzes past and current frames and detects objects on the fly. The baseline code of SORT is written using $300$ lines of Python code and relies on the \verb|scikit-learn| \verb|scipy|, \verb|filterpy|, \verb|numba|, \verb|scikit-image| libraries. 

The evaluation of SORT is done on the Multiple Object Tracking (MOT) benchmark \cite{Mot} data set that has 11 input files. Table \ref{tab:dataset} shows properties of the given dataset.
\begin{table}[htbp]
  \centering
  \caption{DataSet Property}
  \resizebox{0.45\textwidth}{!}
  {
    \begin{tabular}{p{8em}rp{12em}}
    \toprule
    \textbf{Dataset (video)} & {\textbf{\#Frames}} & \multicolumn{1}{p{12em}}{\textbf{Max Tracked Object}} \\
    \midrule
    PETS09-S2L1. & 795   & \multicolumn{1}{p{12em}}{8} \\
    TUD-Campus. & 71    & \multicolumn{1}{p{12em}}{6} \\
    TUD-Stadtmitte. & 179   & \multicolumn{1}{p{12em}}{7} \\
    ETH-Bahnhof. & 1000  & \multicolumn{1}{p{12em}}{9} \\
    ETH-Sunnyday. & 354   & \multicolumn{1}{p{12em}}{8} \\
    ETH-Pedcross2. & 837   & \multicolumn{1}{p{12em}}{9} \\
    KITTI-13. & 340   & \multicolumn{1}{p{12em}}{5} \\
    KITTI-17. & 145   & \multicolumn{1}{p{12em}}{7} \\
    ADL-Rundle-6. & 525   & \multicolumn{1}{p{12em}}{11} \\
    ADL-Rundle-8. & 654   & \multicolumn{1}{p{12em}}{11} \\
    Venice-2. & 600   & \multicolumn{1}{p{12em}}{13} \\
    \bottomrule
    \end{tabular}%
    }
  \label{tab:dataset}%
\end{table}%

\subsection{Kalman Filtering}
\begin{figure}[t]
  \centering
  \includegraphics[scale=0.4]{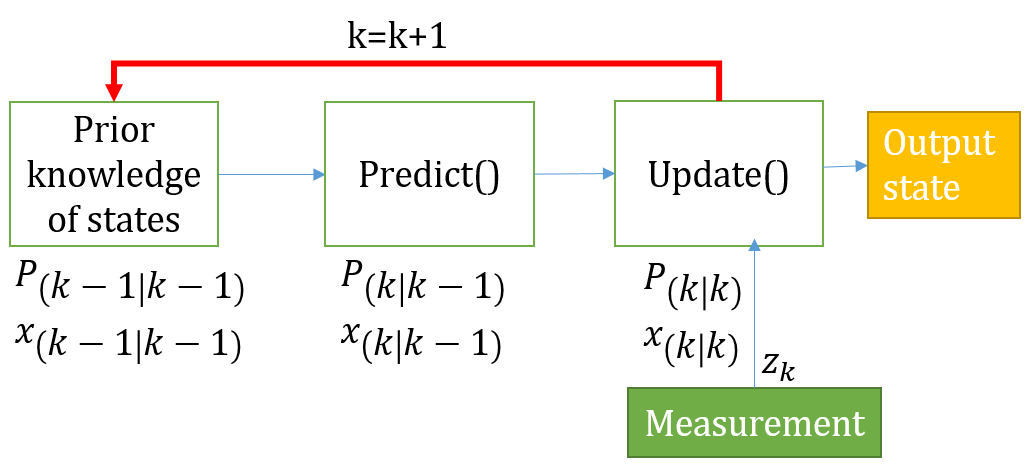} 
  \caption{Kalman Filter workflow.}
  \label{fig:sort:KF}
\end{figure}

Kalman filtering \cite{KalmanFilter0,KF} is an algorithm that uses a series of measurements observed over time, containing statistical noise to produce estimates of unknown variables that tend to be more accurate than those based on a single measurement alone, by estimating a joint probability distribution over the variables for each timeframe. Figure \ref{fig:sort:KF} shows the workflow of a standard Kalman filter. The algorithm works in two steps: prediction and update. In the prediction step, the Kalman filter produces estimates of the current state variables $x(k|k-1)$, along with their uncertainties $P(k|k-1)$. Once the outcome of the next measurement $z_k$ is observed (necessarily corrupted with some amount of error, including random noise), these estimates are updated ($x(k|k), ~P(k|k)$) using a weighted average. It can run in real-time, using only the present input measurements, previously calculated state and a given uncertainty matrix; no additional past information is required. There is a sequential dependency in between the Kalman filter states across consecutive frames of a video sequence which limits parallelism. 

\subsection{Hungarian Algorithm}
The Hungarian method\cite{kuhn1955hungarian,Hungarian} is a combinatorial optimization algorithm that solves the assignment problem in polynomial time. There are two ways to formulate the problem: as a matrix or as a bipartite graph. In SORT, the Hungarian algorithm is used in its matrix format. Given a nonnegative $n \times m$ matrix, where the element in the $i$-th row and $j$-th column represents the cost of assigning the $j$-th predicted object to the $i$-th measured object, the Hungarian algorithm finds an assignment of the prediction to the measurements, such that each prediction is assigned to one measured object and each measured object is assigned one prediction, and the total cost of the assignment is minimum. The algorithm works by subtracting the unique minimum element at each row and column and picking the unique $[i, j]$ entry in a row and column with a $0$ value and then assigns prediction $j$ to measured object $i$. If there are multiple $0$ in a row or column, the process repeats by subtracting a global minimum from all remaining entries. 

\subsection{Kalman Filter and Hungarian Algorithm in SORT}
At a high-level, the Kalman Filter is used to predict the object locations in a video frame, and then the Hungarian algorithm is used to assign the prediction to the measured objects in the frame so that the matching score is maximized. Next, based on the assigned object, the status of the KalmanFilter gets updated and the process is repeated for the next frame. 

\subsection{Challenges}
Internally, KalmanFiltering and Hungarian algorithms manipulate matrices. Kalman filter includes matrix multiplications, dot products, transpose, Inverse, Cholesky factorization and transcendental operations. The Hungarian algorithm involves row-wise/column-wise matrix operations and iterative scaling. Table \ref{tab:sort:kernels} shows a list of matrix kernels and the sizes of matrices used in SORT. Details about those matrices can be found in \cite{filterpy}.
\begin{table*}[htbp]
  \centering
   \caption{Frequently used kernels inside SORT python code.}
  \resizebox{0.8\textwidth}{!}
   {
    \begin{tabular}{ll}
    \toprule
    \rowcolor[rgb]{ .267,  .447,  .769} \textcolor[rgb]{ 1,  1,  1}{\textbf{Kernel}} & \textcolor[rgb]{ 1,  1,  1}{\textbf{Size In/Out}} \\
    \midrule
    \rowcolor[rgb]{ .906,  .906,  .906} \textbf{Matrix-Matrix Multiplication } & H[4][7], P[7][7], Q[7][7], B[7][4] \\
    \textbf{Matrix-Vector Multiplication } & F[7][7], H[4][7], R[4][4], P[7][7], Q[7][7], x[7][1], B[7][4], u[4][1] \\
    \rowcolor[rgb]{ .906,  .906,  .906} \textbf{Matrix-Transpose} & F[7][7], H[4][7] \\
    \textbf{Matrix-Inverse} & S[7][4] \\
    \rowcolor[rgb]{ .906,  .906,  .906} \textbf{Element-wise Matrix-Matrix (add, sub, mul, min) } & Det[12][5], R[4][4], Q[7][7] size varies 1x10 to 13x10 \\
    \textbf{Element-wise Matrix-Vector (add, sub, mul, min) } & Det[12][5], R[4][4], x[7][1], u[4][1], F[7][7], Det size varies 1x10 to 13x10 \\
    \rowcolor[rgb]{ .906,  .906,  .906} \textbf{Element-wise Vector-Vector (add, sub, mul, min) } & x[7][1], z[1][12], u[4][1], size varies 1x10 to 13x10 \\
    \textbf{Matrix-vector manipulation, creation libs} & F[7][7], H[4][7], R[4][4], P[7][7], Q[7][7], x[7][1], B[7][4], u[4][1] \\
    \bottomrule
    \end{tabular}%
    }
  \label{tab:sort:kernels}%
\end{table*}%

The task of object detection in a video sequence is inherently sequential, i.e., we need to process the frames one after the other so that we can use the updated Kalman Filter from the prior frame. However, parallelism exists inside a single frame which varies with the number of objects on that frame. Although it is not challenging to parallelize KalmanFiltering using optimized matrix libraries (for multiplication, Inverse, Cholesky, transpose, dot product, etc), it becomes a challenge to get any runtime benefit, if the input and output matrices are very small. In fact, parallelization slows down the program. In this paper, we address the parallelization challenge of SORT on shared memory multicore architecture.

\section{Object Tracking Algorithm}
In this section, we give a high-level algorithmic explanation of the SORT workload. Algorithm \ref{alg:sort:complete} shows a high level pseudocode for SORT. In case of online tracking, the input video sequence is streamed through the system and each video contains data for multiple frames. 
 
For a stream of video sequences, each frame data is read in a dense matrix of size $N_{R_f} \times N_s$, where $N_{R_f}$ denotes the number of observations (i.e., detected objects) available in the same frame and $N_s$ denotes the number of sensor outputs. For each frame, it calls the update function 
(Line 5-9) which is the ``only timed'' function and takes around $78$\% to $80$\% of the total run time of SORT. 

\begin{algorithm}[h]
\SetAlgoLined
\KwData{Frames, KalmanFilter}
\KwResult{Tracked Object and Updated KalmanFilter}
  \For{each video}
  {
  Create KalmanFilterTracker \;
   \For{each frame in the video}
   {
   Prepare the frame data\;
  //Trk=Update\_frame()\;
  Predict\;
  Assign \;
  Update KalmanFilterTracker \;
  Trk=Updated\_frames\;
  \For {each Trackers in Trk}{
  Write to output file and display.\;
  }
   }
  }
\caption{The Simple Online Object Real-time Tracking (SORT) Algorithm.}
\label{alg:sort:complete}
\end{algorithm}

\begin{figure}[bhtp]
  \centering
  \includegraphics[scale=0.6]{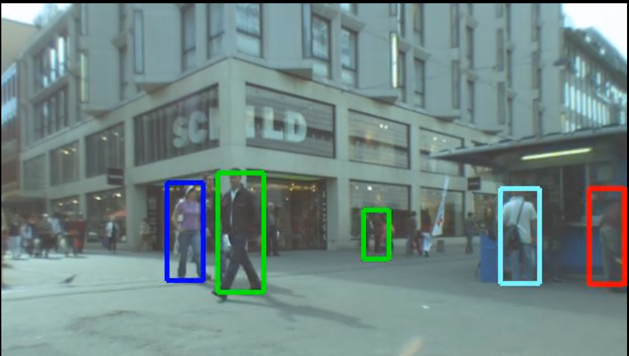}
  \includegraphics[scale=0.6]{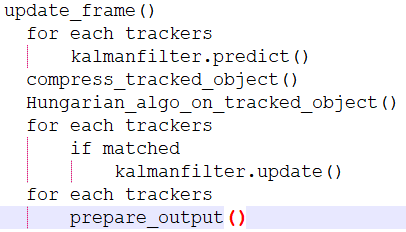} 
  \caption{Example of object tracking on a video frame and pseudocode of the Update function.}
  \label{fig:sort:update}
\end{figure}

Figure \ref{fig:sort:update} shows the pseudocode of the {{Update}} function. It first uses the existing Kalman Filter Trackers to predict potential objects' positions on the frame and stores them into a Trackers array, Trk. It then compresses the Trackers array by removing all rows with invalid data (if any). Next, it tries to assign (or match) the predicted objects with the actual objects on the frame using the Hungarian algorithm \cite{kuhn1955hungarian}. From this step, three lists are created: 1) matched predictions, 2) unmatched predictions and 3) unmatched detections on the frame. Next, the current Kalman filters' internal states associated with the matched predictions are updated using the assigned detections. For each unmatched object, a new Kalman Filter tracker is created using the unmatched object as a seed. Finally, it prepares the output from the matched detections and trackers and at the same time discards outdated trackers (if any). 

A simple timing model for SORT can be written as follows: $T_{frame} = a~ T_{Prediction} + b~ T_{Assignement} + c~ T_{Update} + d~ T_{Output prep + Trackers update}$, where, $a$, $b$, $c$ and $d$ are multiplicative parameters. 
\section{Workload Analysis}
We use the server platform {{Intel(R) Xeon(R) Gold 6140 CPU @ 2.30GHz, CPU 1800 MHz, with 25MB shared victim L3 cache}} codenamed Skylake {(SKX)} for our initial workload analysis on the entire application.  
\begin{table}[htbp]
   \caption{Measured hardware performance counters for Object Detection (TLB = translation lookaside buffer, LLC = last level cache, MPKI = misses per thousand instructions, BW = memory bandwidth)}
\resizebox{0.5\textwidth}{!}
{
   \centering
   \begin{tabular}{|c|c|c|c|c|c|}
    \toprule
    \textbf{Instructions} & \textbf{Time (s)} & \textbf{IPC} & \textbf{TLB MPKI} & \textbf{LLC MPKI} & \textbf{BW usage} \\
    \midrule
    4.755E+10    & 10           & 2.21         & 0.136        & 0.059        & 0.015\% \\
    \bottomrule
    \end{tabular}%
    }
   \label{table:obj_detect_metrics_scale}
\end{table}

While the SORT implementation is single-threaded, it uses internally multi-threaded libraries. Table \ref{table:obj_detect_metrics_scale} shows values for performance counters while running the python program. The table shows a low Instructions per cycle (IPC) value, especially since this is an aggregated IPC over all threads (SKX has a peak IPC of $4$ per core). The TLB and LLC Miss Per Kilo Instructions (MPKI) is low as well, revealing these components are not limiting performance. Memory bandwidth usage is very low, so the application is not memory bandwidth bound. Neither is it memory latency bound since the LLC MPKI is very low. We assume performance is limited by the high amount of time that is spent in system calls and thread synchronization. 
 
\begin{figure*}[htpb]
  \centering
  \includegraphics[width=0.8\textwidth]{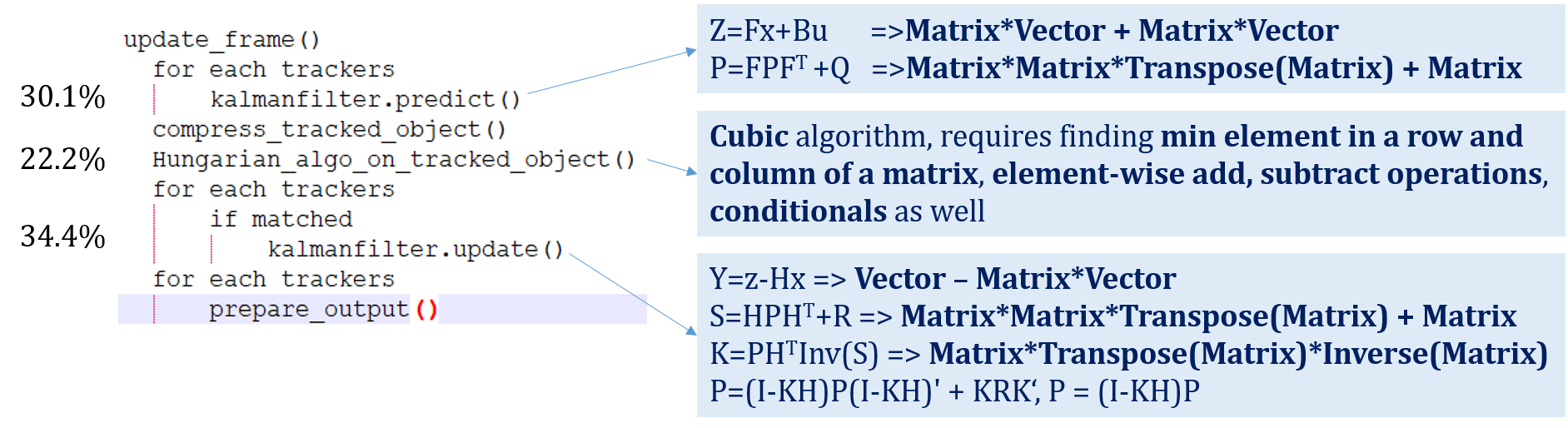} 
  \caption{Cprofile stats of the Update function and details for matrix operations performed inside each.}
  \label{fig:sort:update_cprofile}
\end{figure*}
 
We used \verb|cprofile| to profile the {{Update}} function (Figure \ref{fig:sort:update_cprofile}). About $30$\% of the time is spent in the prediction phase, $22.2$\% time is spent in the assignment phase (Hungarian algorithm) and $34.4$\% time is spent in the update function. The remaining part is spent in preparing the output array. The Kalman ``predict'' and ``update'' functions internally use a number of common matrix kernels as shown in Table \ref{tab:sort:update_AI}. Python parallel BLAS and LAPPAC modules used to handle most of these matrix and vector operations. The Hungarian algorithm comes from the linear assignment python library.

\begin{table*}[htbp]
    \caption{Algorithm steps, compute kernels and arithmetic intensity (AI) of each step.}
  \centering
   \resizebox{0.8\textwidth}{!}
   {
    \begin{tabular}{lp{34.4em}cc}
    \toprule
    \rowcolor[rgb]{ .267,  .447,  .769} \multicolumn{1}{c}{\textcolor[rgb]{ 1,  1,  1}{\textbf{Step1}}} & \multicolumn{1}{c}{\textcolor[rgb]{ 1,  1,  1}{\textbf{Compute Kernels}}} & \textcolor[rgb]{ 1,  1,  1}{\textbf{\% of time}} & \textcolor[rgb]{ 1,  1,  1}{\textbf{AI}} \\
    \midrule
    \rowcolor[rgb]{ .906,  .906,  .906} \textbf{6.2.predict} & DGEMV + DGEMM + Transpose + Mat-Mat + Vect-Vect + sqrt & 30 & 2.4 \\
    \textbf{6.3 assignment} &  Mat\_copy + Vec\_reset + f($N^2{_{R_{f}}} \times N^2{_{{T}_f}} + N_{R_{f}} \times N_{T_{f}} \times N_s$) & 22.2  & 1.5 \\
    \rowcolor[rgb]{ .906,  .906,  .906} \textbf{6.4 update} & Transpose + DGEMM + DGEMM +  Mat-Mat + cholesky/Inv + DGEMM + DGEMM + DGEMV+ Vec-Vec-add +  DGEMV + Vec-Vec-sub + DGEMM + Mat\_negate + Mat\_add\_eye + DGEMM + Vec-Vec + div & 34.3  & 18 \\
    \textbf{6.6 create new} & Scalar*Matrix & 3.1   & 0.1 \\
    \rowcolor[rgb]{ .906,  .906,  .906} \textbf{6.7 prepare output} & $N^2{_{R_{f}}}\times N_s +2N^2{_{{T}_f}} \times N_s$ & 9.9   & 1 \\
    \bottomrule
    \end{tabular}%
    }
  \label{tab:sort:update_AI}%
\end{table*}%

\section{Our Version of SORT}
Our C program for SORT follows algorithmic steps of the original python code. We utilized open-source Kalman Filter \cite{KF} and Hungarian algorithm \cite{Hungarian} implementations and optimized them as needed. Table \ref{tab:sort:update_AI} shows a tabular representation of the algorithmic steps, arithmetic intensity (flops per byte) of each step and \% of the time taken. Here, $N_{T_f}$= Number of trackers used for frame $f$. Table \ref{tab:summary} shows a summary of the performance of our code running on 1 core vs. the original code running on all cores of a given machine. Our sequential code obtains $45$ to $106\times$ speedup overall. Intel Vtune profiler shows that in C implementation, the matmul takes the most time which appears in predict and update functions of Kalman filter.

\begin{table}[htbp]
\caption{Speedup wrt. the original code.}
   \centering
   \resizebox{0.5\textwidth}{!}
   {
    \begin{tabular}{p{14.3em}|cc|c}
    \toprule
    \multirow{2}[4]{*}{Server Machine} & \multicolumn{2}{p{10.3em}|}{Time (s)} & \multicolumn{1}{c}{\multirow{2}[4]{*}{Speedup}} \\
\cmidrule{2-3}    \multicolumn{1}{c|}{} & \multicolumn{1}{p{3.75em}}{C (ours)} & \multicolumn{1}{p{6.55em}|}{Python (orig.)} &  \\
    \midrule
    Xeon(R) 6140 CPU @ 2.30GHz & 0.12  & 5.4   & 45 \\
    \midrule
    Xeon(R) 8280 CPU @ 2.70GHz & 0.074 & 7.9   & 106.76 \\
    \bottomrule
    \end{tabular}%
    }
  \label{tab:summary}%
\end{table}%

\section{Strong-, Weak- and Throughput-scaling}
In this paper, by strong-scaling, we mean parallelizing the task of processing a single video sequence. Since there is sequential dependency across frames in a video sequence in terms of states of the Kalman Filter, strong-scaling requires that the object detection inside a single frame is done in parallel and each frame is processed using $p$ cores where $p \in [1, 18, 36, 72]$. 

In the case of weak-scaling, instead of parallelizing the object detection inside a single frame, parallelization happens across input files (entire video sequence). In weak-scaling, we use $1$ core to process $1$ video file independently. Thus, $11$ video files are processed by $11$ cores in parallel. This version should stop scaling after $11$ cores. 

In the case of throughput scaling, we run $p$ independent sequential SORT executables on $p$ cores and each of those programs processes $k$ independent video files in parallel, i.e., $pk$ files in total. The difference between weak-scaling and throughput scaling is: weak-scaling runs $1$ executable using $p$ cores, and throughput-scaling runs $p$ executables each using $1$ core. In throughput scaling, each of the core gets a completely independent fraction of shared resources (cache, memory, bandwidth) whereas, for weak-scaling, threads can share common data and resources.

\begin{table}[htbp]
\caption{Strong-, Weak- and Throughput-scaling}
 \centering
 \resizebox{\linewidth}{!} 
 {
    \begin{tabular}{|r|r|r|r|r|r|}
    \midrule
    \multicolumn{1}{|l|}{Cores} & \multicolumn{1}{l|}{files} & \multicolumn{1}{l|}{frames} & \multicolumn{1}{l|}{Strong} & \multicolumn{1}{l|}{Weak} & \multicolumn{1}{l|}{Throughput } \\
     \midrule
    1     & 11    & 5500  & 37415 & 45082 & 47573 \\
    18    & 11    & 5500  & 24663.7 & 34810.1 & 37450 \\
    36    & 11    & 5500  & 23404.3 & 37162.2 & 37489 \\
    72    & 11    & 5500  & 19503.5 & 31976.7 & 38400 \\
    \bottomrule
    \end{tabular}%
    }
  \label{tab:Xeonscaling}%
\end{table}%

\begin{figure}[htbp]
  \centering
  \includegraphics[width=0.3\textwidth]{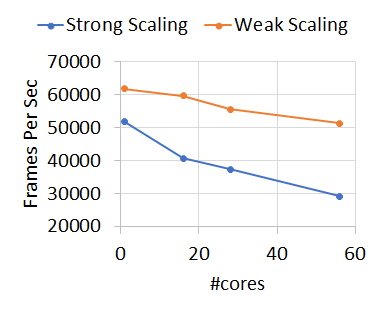} 
  \caption{Strong Scaling vs Weak Scaling.}
  \label{fig:sort:scaling}
\end{figure}

Since the single-core run takes an only fraction of a second, we expect not to see any benefit of strong scaling which was indeed the case. When we parallelized the code to process each frame with multiple OpenMP threads, the overhead of threading turned out to be larger than the benefit due to the limited parallelism. It slowed down the program instead of making it faster. Table \ref{tab:Xeonscaling} shows impact of three different types of scaling on the SKX machine in terms of Frames Per Sec (FPS) processed. The best single-core FPS is $47$k. The best FPS for OpenMP parallel code with strong scaling is $37$k. In strong scaling, performance drops with the increase the in number of cores, because the matrices are too tiny to produce enough work. For throughput scaling, it was able to sustain $37$k FPS as before. We replicated the input files $7$ times and re-ran the weak and strong scaling on Intel(R) Xeon(R) Platinum 8280 CPU machine which is advanced than SKX. Figure \ref{fig:sort:scaling} shows the comparison result. This also shows that weak-scaling works better than strong scaling for SORT. 

\section{Conclusion}
 The Simple Online Real-time Tracking (SORT) is an interesting workload with much real-world significance. In this paper, we parallelize SORT using OpenMP and show that due to the nature of dataset (i.e., small matrices) weak-scaling or throughput scaling performs better than strong scaling. The conclusion might change if the dataset properties change. Our parallel implementation is more than $50\times$ faster than its original implementation. We plan to make our code open source.

\bibliographystyle{IEEEtran}
\bibliography{IEEEabrv,obj-detect}
\end{document}